\documentclass[aps,showpacs]{revtex4}
\usepackage{psfig}

\def\order#1{{\cal O}\!\left(#1\right)}
\newcommand{\ba}{\begin{eqnarray}}
\newcommand{\ea}{\end{eqnarray}}
\newcommand{\be}{\begin{equation}}
\newcommand{\ee}{\end{equation}}

\begin{document}

\title{
%
%
\[ \vspace{-2cm} \]
\noindent\hfill\hbox{\rm  } \vskip 1pt \noindent\hfill\hbox{\rm
Alberta Thy 16-05} \vskip 10pt
%
 Enhancement of the hadronic $b$ quark decays\footnote{This research was supported
 in part by the EU grant MTKD-CT-2004-510126,
in partnership with the CERN Physics Department;  by the Science and
Engineering Research Canada; and by Alberta Ingenuity.}
}

\author{Andrzej Czarnecki}
\affiliation{Department of Physics, University of Alberta\\
Edmonton, AB\ \  T6G 2J1, Canada
\\ and
\\
Institute of Nuclear Physics, Academy of Sciences,\\
ul.~Radzikowskiego 152,
31-342 Krak\'ow, Poland}

\author{Maciej \'Slusarczyk}
\affiliation{
Department of Physics, University of Alberta\\
Edmonton, AB\ \  T6G 2J1, Canada}

\author{Fyodor Tkachov}
\affiliation{
Institute for Nuclear Research, Russian Academy of Sciences\\
Moscow, 117312, Russian Federation}

\begin{abstract}
A class of previously unknown strong-interaction
corrections is found to enhance the rate of non-leptonic decays of the $b$
quark by 5-8 percent. This effect decreases the predicted fraction of
semi-leptonic decays and brings it into fair agreement with
experimental results.  As well as solving a long-standing puzzle
of measurements disagreeing with the Standard Model prediction,
our work suggests a way for future precise studies of non-leptonic
$b$ quark decays and their application to searching for ``new
physics."
\end{abstract}


\maketitle

The $b$ quark is a relatively long-lived particle.  As the lighter
of the third-generation quarks, it can decay only into quarks from
other generations, to which its couplings are suppressed in the
Standard Model.  For more than a decade, theoretical predictions
disagreed with measurements of the relative probabilities of
semi-leptonic and non-leptonic $b$ quark decays.  It has been
speculated that  this discrepancy may be due to some exotic ``new
physics" mechanism, not affected by the Standard Model suppression \cite{Bigi:1993fm}.

\begin{figure}[htb]
\begin{tabular}{c@{\hspace*{15mm}}c@{\hspace*{15mm}}c}
\psfig{figure=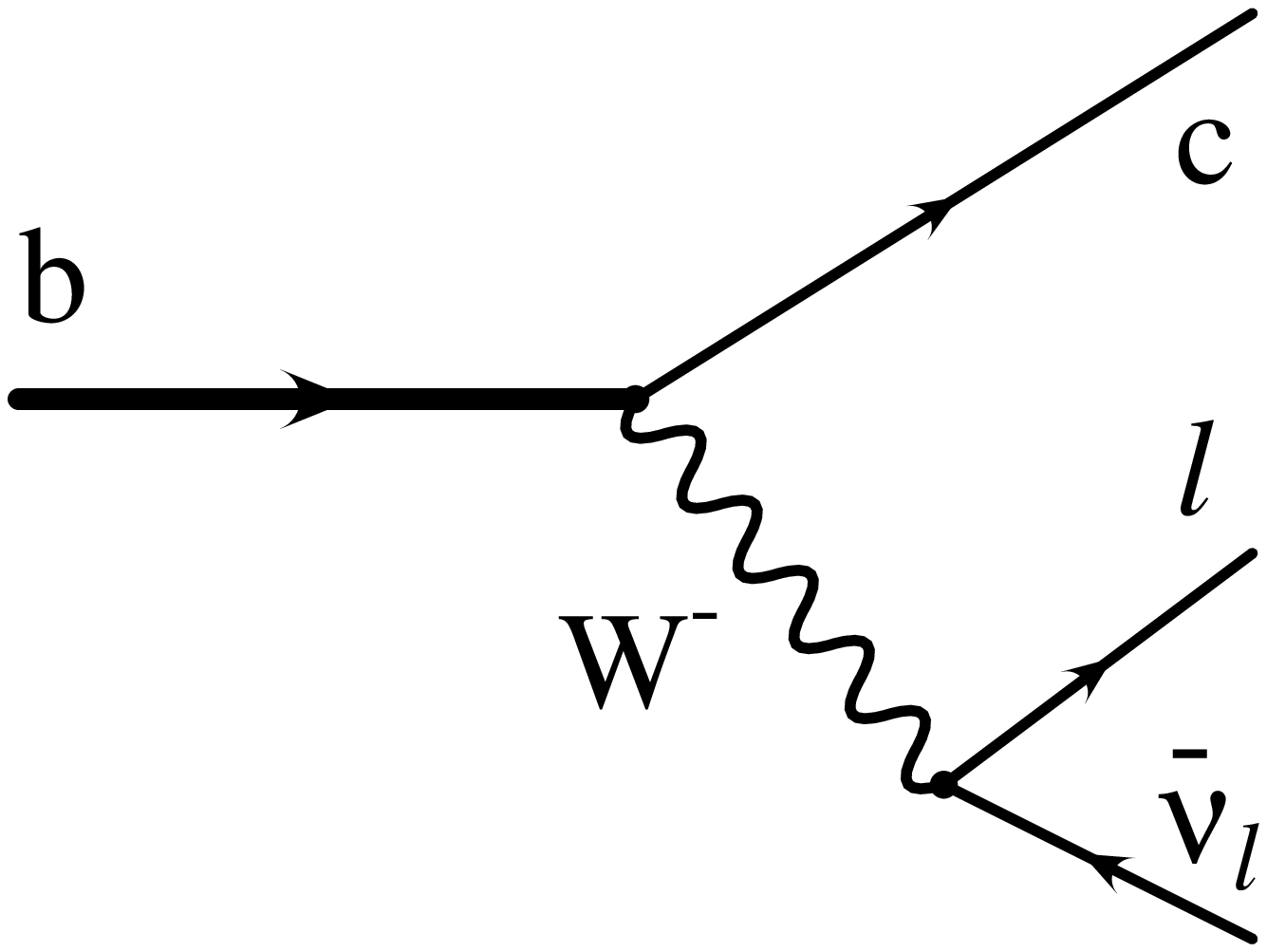,width=40mm} &
\psfig{figure=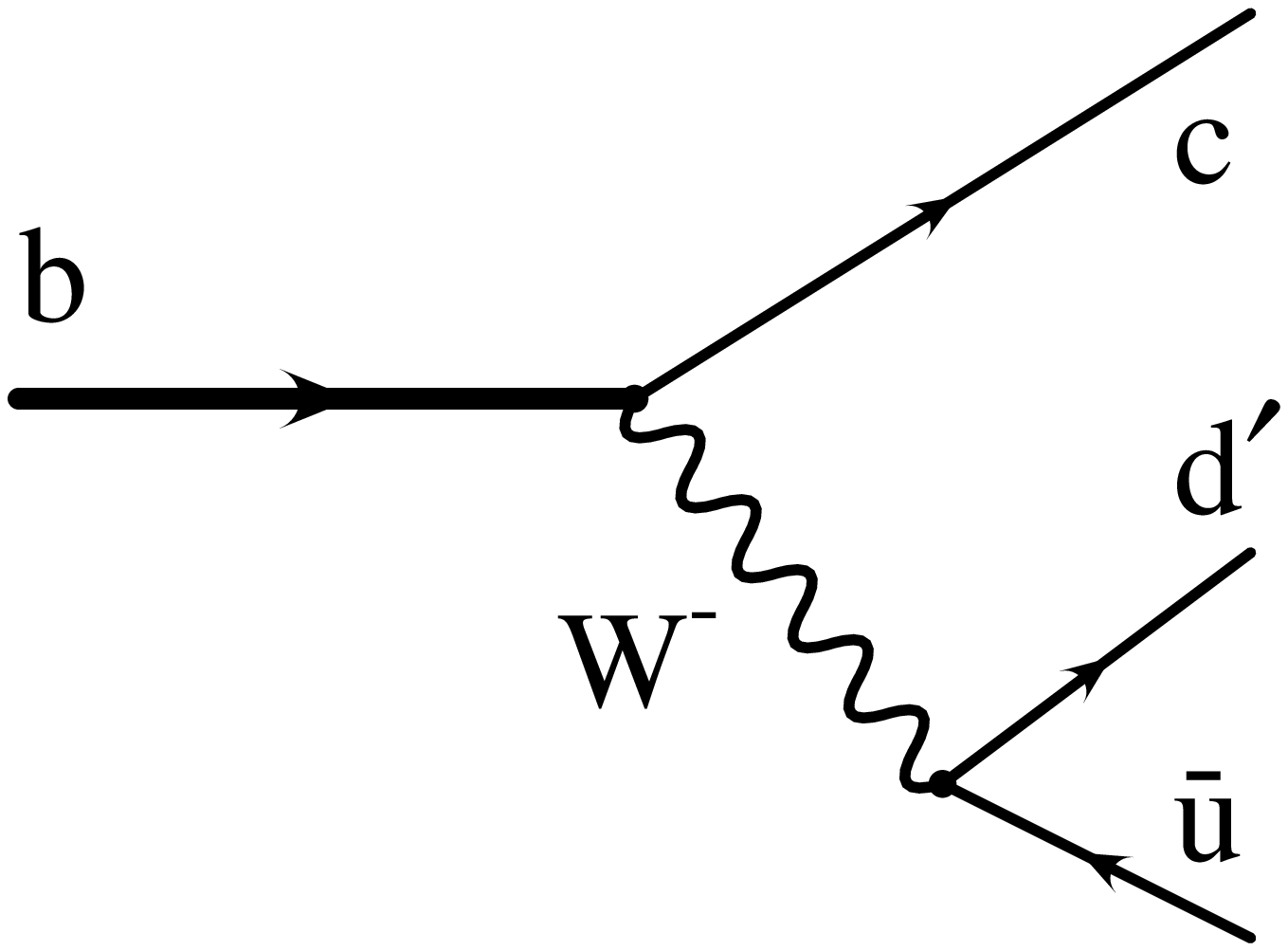,width=40mm} &
\psfig{figure=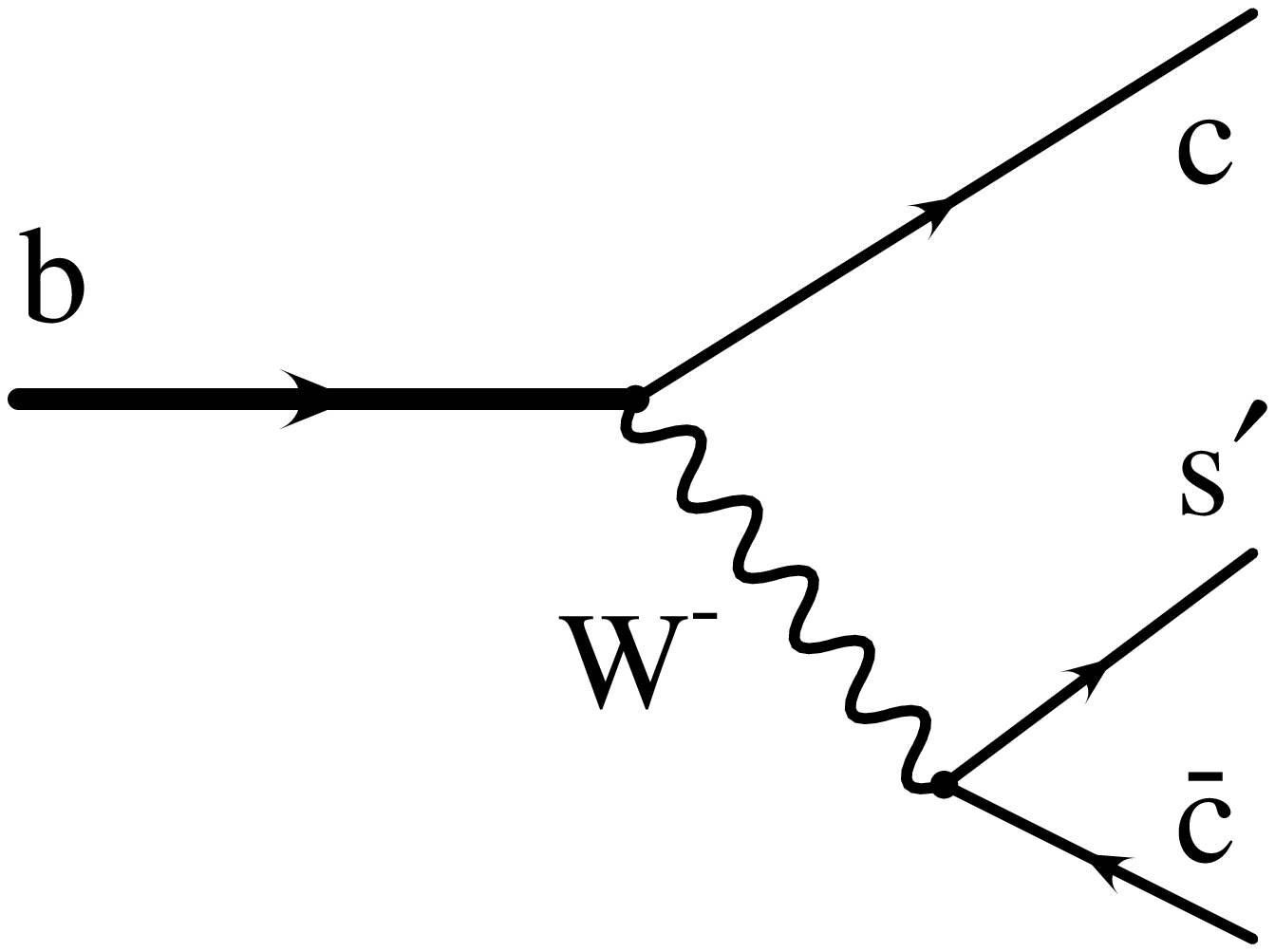,width=40mm}
\\
(a) & (b) & (c)
\end{tabular}
\caption{Main decay channels of the quark $b$:  (a) semi-leptonic, (b) non-leptonic with light quarks,
(c) non-leptonic with an extra charm quark.  $d'$ and $s'$ denote the approximate flavor eigenstates: combinations
of $d$ and $s$ quarks which couple to $u$ and $c$ in weak decays.  }
\label{fig:mainDecays}
\end{figure}

The main decays of the  $b$  quark are the semi-leptonic channels $b\to c l \bar \nu_l$
(with $l$ being an electron, a muon, or a $\tau$ lepton), and the hadronic channels
$b \to c \bar u d'$ and $b \to c \bar c s'$, as shown in Fig.~\ref{fig:mainDecays} ($d'$ and $s'$ denote
approximate flavor eigenstates coupling to $u$ and $c$, respectively).
In addition, $b$ can also decay into an $u$ quark, $b\to u X$, or through one of the radiative
channels such as $b\to s\gamma$, which arise due to quantum loop effects. These decays are rare and
will be ignored in this paper.
The probability of a semi-leptonic decay such as $b\to c e \bar \nu_e$ is known as the
semi-leptonic branching ratio $B_{\rm SL}$.  It can be expressed in terms of the
partial decay rates $\Gamma$,
\ba
B_{\rm SL} \left( b\to c e \bar \nu_e \right) = {\Gamma\left( b\to c e \bar \nu_e \right)
\over \sum_{l=e,\mu,\tau} \Gamma\left( b\to c l \bar \nu_l \right)
+  \Gamma\left( b \to c \bar u d'\right)+  \Gamma\left( b \to c \bar cs'\right)+\Gamma_{\rm rare}}
\label{def}
\ea
 The latest published measurement \cite{Mahmood:2004kq}
found
\ba
B_{\rm SL}^{\rm exp}(B\to Xe^+ \nu_e) =(10.91\pm 0.26)\%.
\label{exp}
\ea
Note that this experimental result refers to a $B$ meson rather than to the $b$ quark; the difference
between their branching ratios is expected to be small, as is discussed below.

Theoretical predictions of $B_{\rm SL}$ have been significantly
higher than the measured values.  Ref.~\cite{Bigi:1993fm} thoroughly
analysed possible non-perturbative effects and all perturbative ones
known at that time \cite{Altarelli:1991dx},
and concluded that
$B_{\rm SL}$ should be not less than 12.5 percent.  An important point
made in that study was that the non-perturbative effects are small,
in particular those that differentiate between the free quark and
the meson decay rate.

Subsequently it was found \cite{Voloshin:1994sn,Bagan:1994qw} that
perturbative corrections enhance the decay channel $ b \to c \bar
cs'$ by about 30 percent, due to the slowness of the massive charm
quarks in the final state. Such enhancement of the hadronic rate
decreases the theoretical lower limit on $B_{\rm SL}$ by about one
percentage point. At the time of its publication, such
explanation was controversial \cite{Falk:1994hm}, since it would
increase the average number $n_c$ of charm quarks  produced in $b$
decays, which seemed to contradict the data  (see also a discussion
in  \cite{Neubert:1996we} which in addition studied spectator effects).
Very recently a new
measurement \cite{Aubert:2004nb} found a larger value of $n_c$, so
the enhancement of $ b \to c \bar cs'$  is no longer out of
question.

In this paper
the value taken from a recent review \cite{Voloshin:2000zc} is adopted as a reference point
for the theoretical prediction,
\ba
B_{\rm SL}^{\rm theory}(B) > 11.5\%.
\label{theory}
\ea
This number is based on the analysis of perturbative and
non-perturbative effects of \cite{Bigi:1993fm} and an enhancement of
the $ b \to c \bar cs'$ channel. Despite the enhancement of
$\Gamma(b\to c\bar c s')$, it still exceeds the experimental value,
Eq.~(\ref{exp}), by about 2.3 standard deviations (2.3$\sigma$).
Assuming that the experimental number remains constant, what effects
can change the theoretical limit and bring it into agreement with
observations?  First of all, whatever affects the semi-leptonic rate
in Eq.~(\ref{def}), has a very similar impact on the non-leptonic
rate and thus cancels in the ratio.
Thus, the most important effects controlling $B_{\rm SL}$ are the
corrections to the non-leptonic decays.

In order to  bring theory and experiment into agreement within
one sigma, the theoretical value of $B_{\rm SL}$ should decrease by
a third of a percentage point. To get a rough estimate of the
required change of the non-leptonic rate, consider a limit in which
all leptons and light quarks are massless, and neglect all
interactions among decay products.  Then Eq.~(\ref{def}) gives
$B_{\rm SL} \simeq {1\over 3+6} = 11.1\%.$  The numbers in the
denominator account for the three semi-leptonic channels and the two
non-leptonic channels; the widths of the latter are enhanced by a
color factor of three.  In order to lower this branching ratio by
one-third of a percentage point, the non-leptonic
width should increase by 4.6 percent.  In this paper,  a class of
strong-interaction effects --- that previously could not be
evaluated --- are found to provide just such an enhancement.

\begin{figure}[htb]
\begin{tabular}{c@{\hspace*{5mm}}c@{\hspace*{5mm}}c}
\psfig{figure=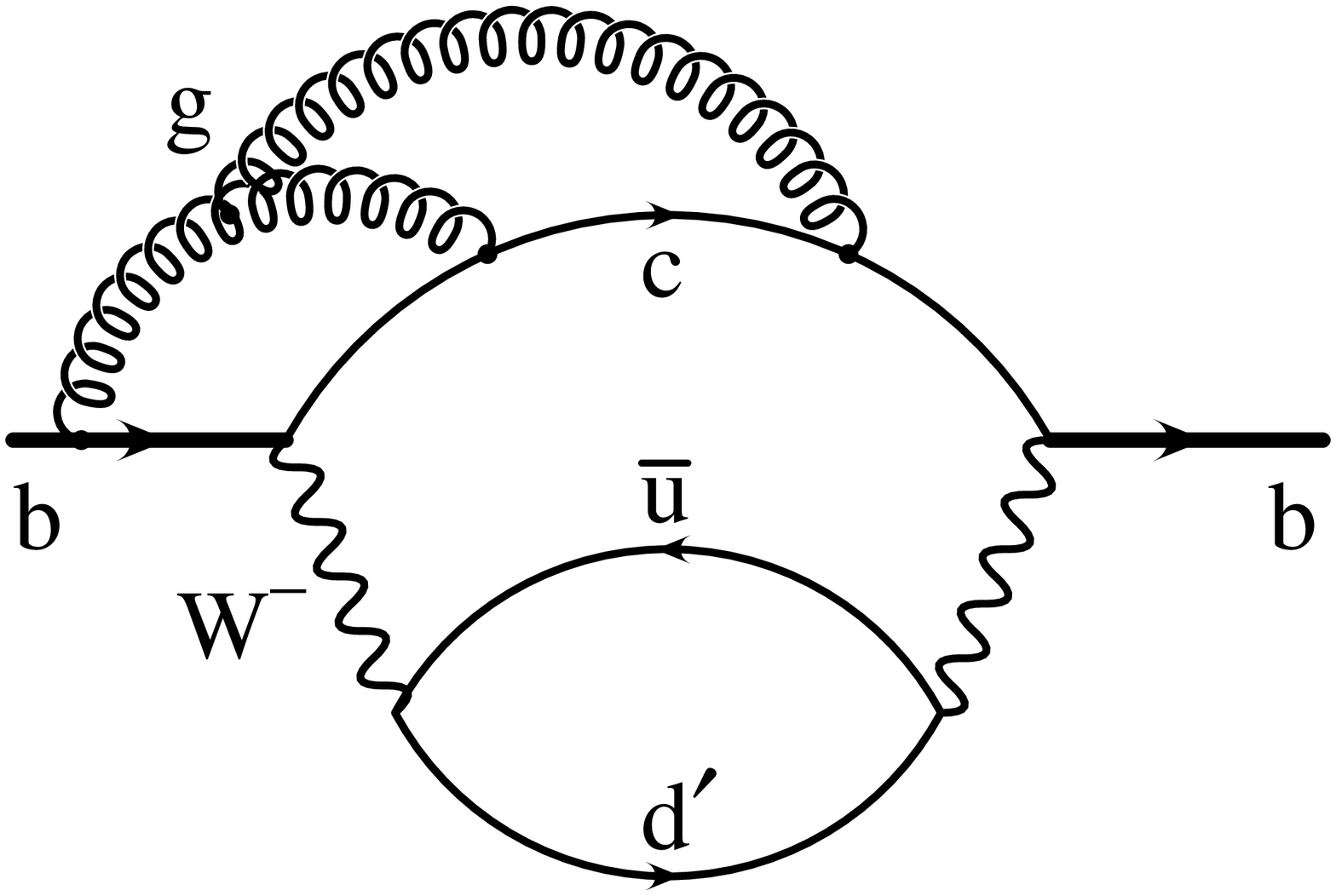,width=45mm} &
\psfig{figure=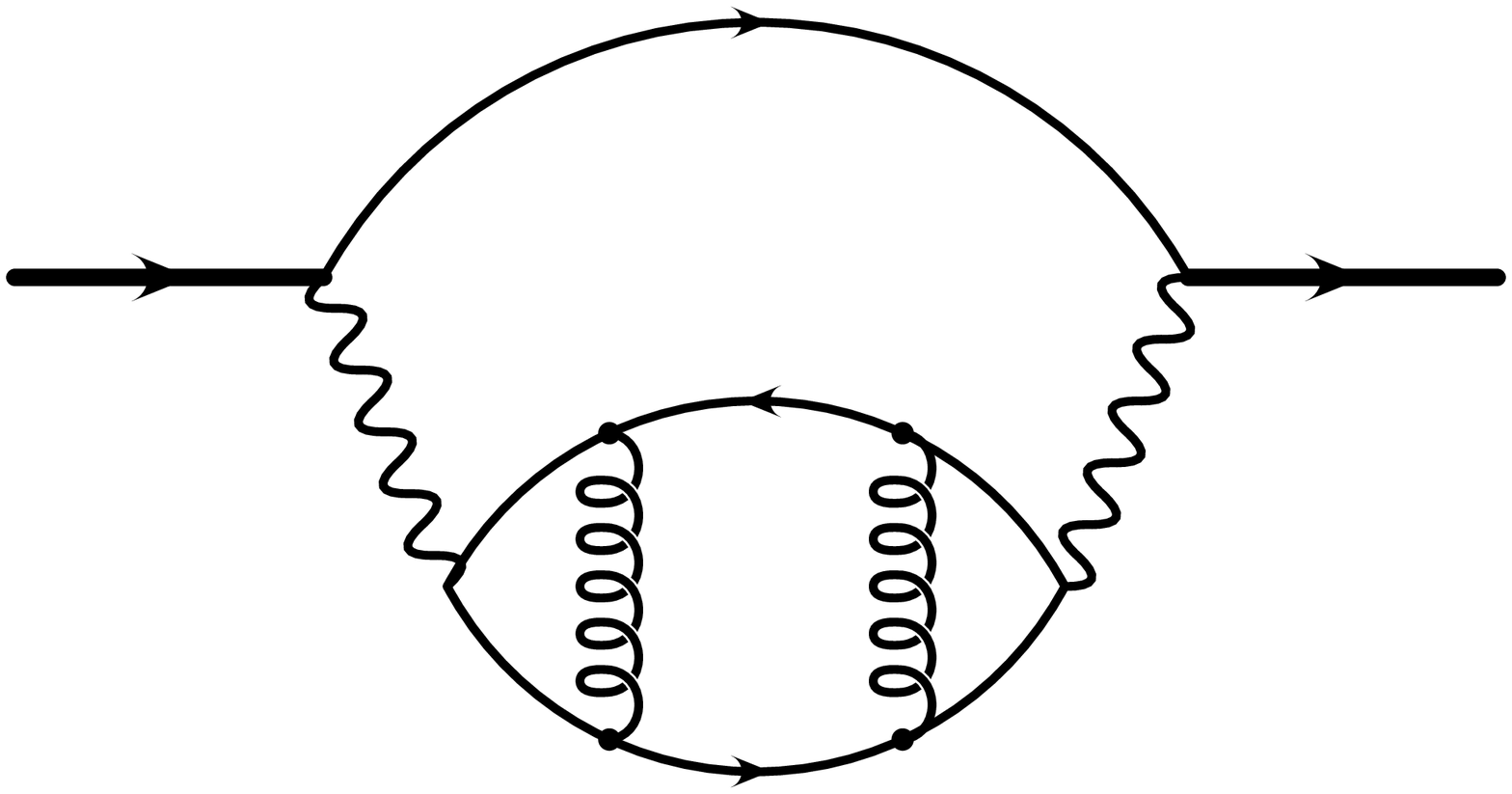,width=45mm} &
\psfig{figure=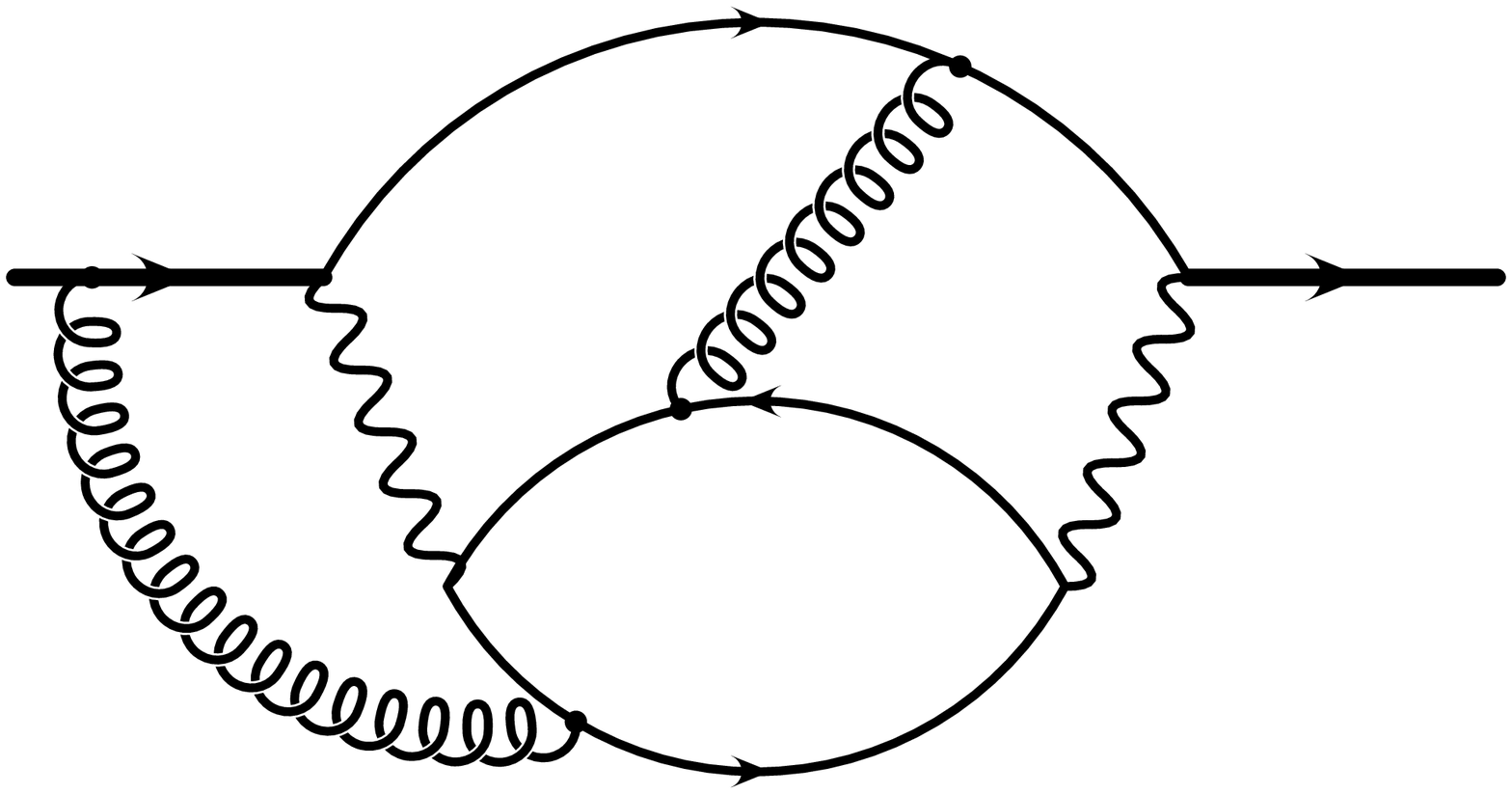,width=45mm}
\\
(a) & (b) & (c)
\end{tabular}
\caption{Examples of the three types of QCD corrections to the {\em
squares} of the non-leptonic
  decay amplitudes:  (a) corrections on the heavy quark line, (b) on the light quark
  line, (c) between the lines. The solid and wavy lines corresponds to the same fields as in Fig.~\protect\ref{fig:mainDecays}(b),
  and the springs denote gluons.}
\label{fig:tri}
\end{figure}

Since it has been established that the size of $B_{\rm SL}$ is
controlled by the perturbative QCD corrections \cite{Bigi:1993fm},
we briefly review what diagrams describe those effects.  To this end
it is convenient to consider the imaginary parts of the forward
scattering amplitudes, such as those in Fig.~\ref{fig:tri}.  There are
two separate quark lines: that continuing from the
incoming $b$ quark, and a closed loop containing only lighter
quarks.  An analysis of $B_{\rm SL}$  is  simplified by the almost
exact cancellation of corrections due to gluon exchanges on the $bc$
line, like the diagram in Fig.~\ref{fig:tri}(a).  These corrections
are common to semi- and non-leptonic corrections and cancel in the
ratio (\ref{def}) (up to residual effects due to small phase space
differences). Exchanges of gluons between the light quarks, such as
Fig.~\ref{fig:tri}(b), have already been studied in great detail in
the context of $\tau$ lepton decays. Thus, the only class specific
to our problem is the interaction between both quark lines,
Fig.~\ref{fig:tri}(c).

Since the light quark pair is produced as a color singlet, at least
two gluons have to be exchanged in an interaction between the two
quark lines. Such corrections arise only in the second order in the
strong coupling constant $\alpha_s$, and they are similar to
electrodynamic (QED) interactions (no diagrams with non-abelian
three-gluon vertices contribute at this order).
Fig.~\ref{fig:tri}(c) shows one of the twelve types of diagrams,
which must be calculated.

The evaluation of diagrams of type (c) has been considered a
daunting challenge.  Indeed, they are four-loop diagrams that depend
on two masses, $m_W$ and $m_b$.  We first explain how the hierarchy
of mass scales, $m_b \ll m_W$, somewhat simplifies the task and then
describe how  the four-loop diagrams are computed.

Since the $b$-quark is much lighter than the $W$ boson, only
the leading term in the expansion in $m_b/m_W$ is needed.  In the language
inspired by
the asymptotic operation \cite{Tkachev:1994gz},
two characteristic virtualities $q^2$ for each of the two gluons should be
considered:
their scales are $m_W^2$ and $m_b^2$.  The hard gluons ($q^2 \sim
m_W^2$) modify the Wilson coefficients of the four-fermion effective
operators.  The soft ones ($q^2 \sim m_b^2$) correct their matrix
elements.

The soft and hard effects are not separately finite, hence the final
result contains logarithms of the ratio of the two scales, $\ln^2
{m_W\over m_b}$ and $\ln{m_W\over m_b}$.  In the past, these
logarithmic terms were evaluated and even summed to all orders using the
renormalization group equation \cite{Bagan:1994zd}.  However, the numerical value of the
logarithm is not very large, $\ln{m_W\over m_b}\simeq 2.8$, so the
$L^2 $ and $L$ terms may not be sufficient for a reliable prediction
of the non-leptonic decay rate.  The importance of finding the
non-logarithmic part, provided in
this paper, has been stressed repeatedly.

\begin{figure}[htb]
\begin{tabular}{c}
\psfig{figure=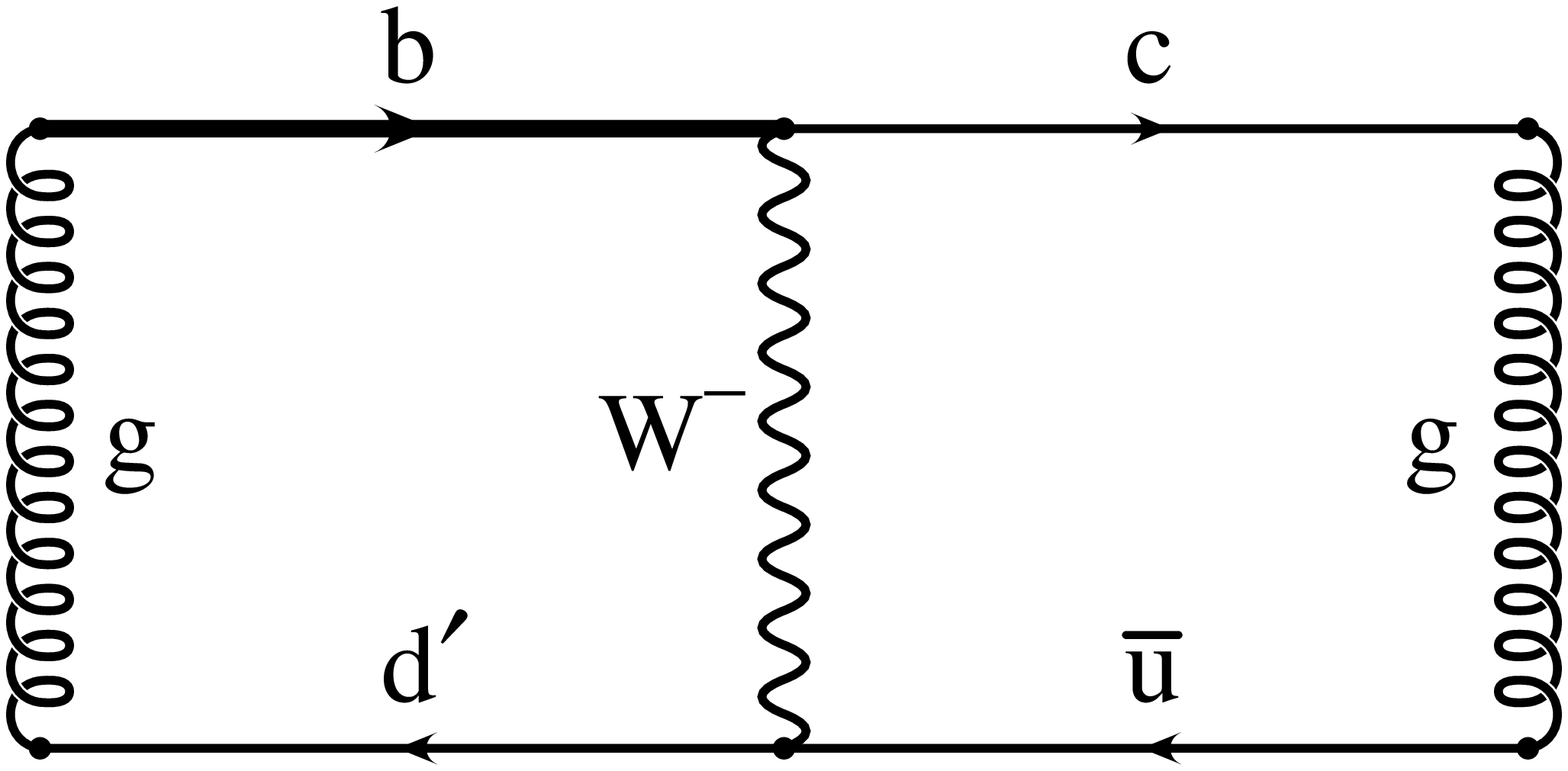,width=40mm} \hspace*{5mm} \raisebox{7mm}{\psfig{figure=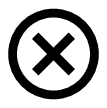,width=3.5mm}}
 \hspace*{5mm}  \psfig{figure=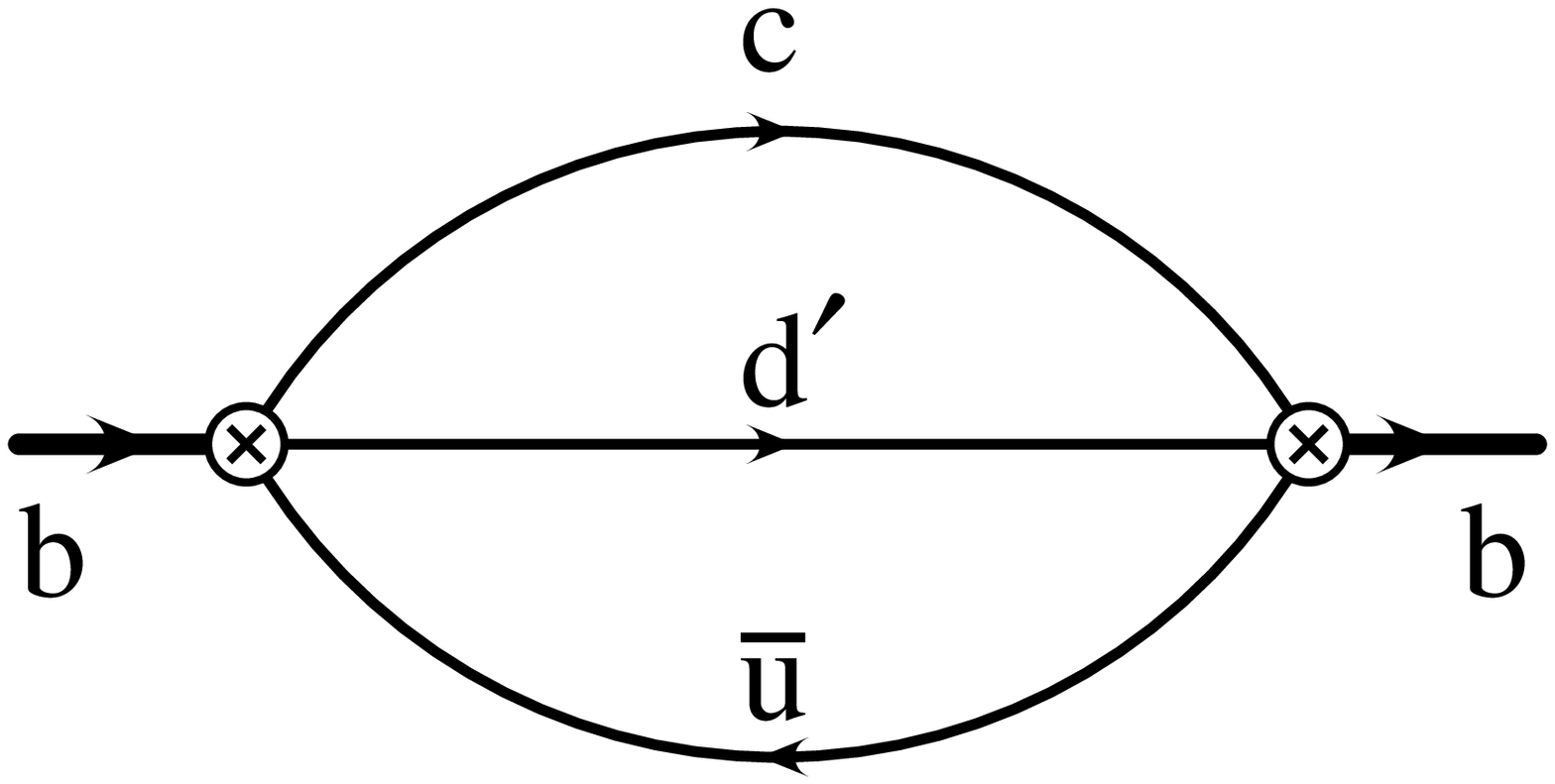,width=40mm} \\[2mm] (a)
\\[4mm]
\psfig{figure=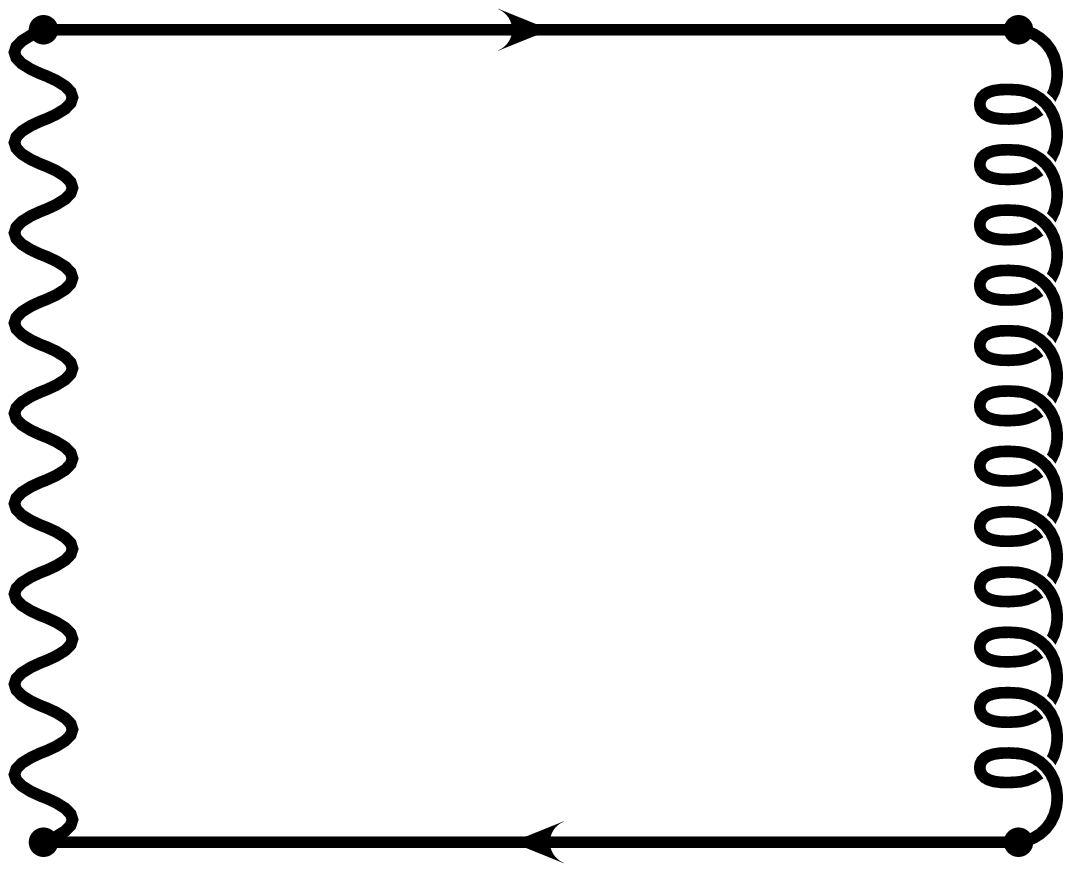,width=22mm} \hspace*{5mm} \raisebox{7mm}{\psfig{figure=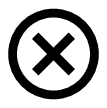,width=3.5mm}}
 \hspace*{5mm}  \psfig{figure=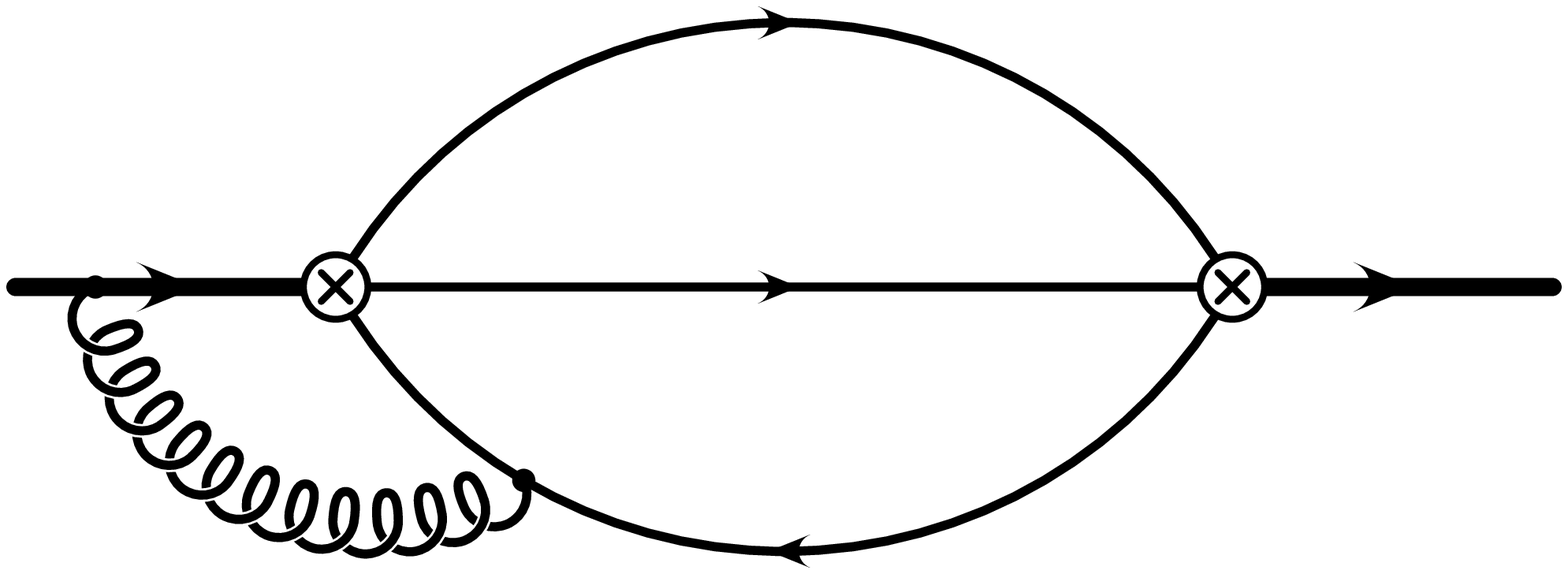,width=45mm} \\[2mm] (b)
\\[5mm]
\psfig{figure=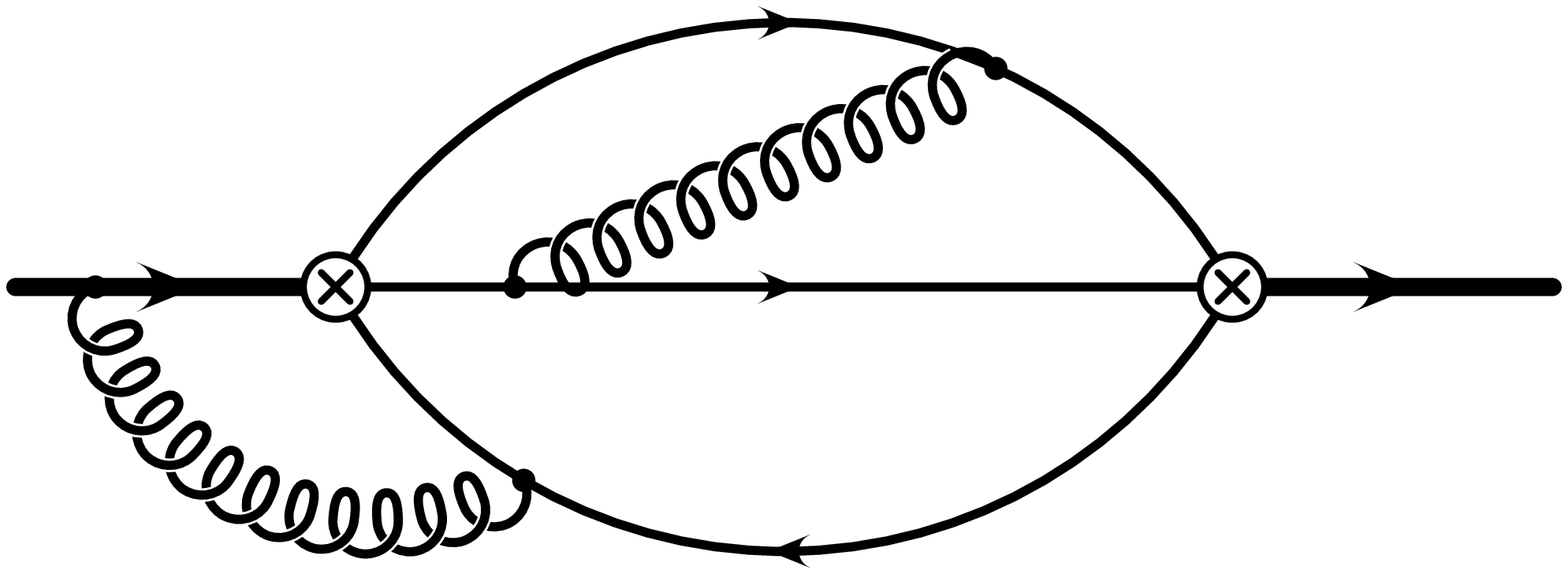,width=48mm}
\\[2mm] (c)
\end{tabular}
\caption{Examples of factorized contributions to the four-loop diagram of Fig.~\protect\ref{fig:tri}(c) from various regions
of virtuality of the two gluons: (a) hard-hard, (b) hard-soft, (c) soft-soft. In (a,b) there is also, not shown,
a second  $W$ boson propagator, which  reduces to an overall factor $1/m_W^2$.  In (c) there are two such factors.
The circles and crosses in the soft subgraphs indicate places from which hard  subgraphs have been taken. }
\label{fig:factors}
\end{figure}

Contributions in which at least one gluon is hard are relatively
easy to evaluate.  The hard momentum flows in a closed loop (since
all external momenta are soft), and this closed subgraph can be
shrunk to a point in the remaining soft part
of the diagram.  Examples of how this is done in practice are shown
in Fig.~\ref{fig:factors}(a,b). The hard subgraphs have no external
legs and are easy to compute.  In hard-soft diagrams
three-loop soft subdiagrams appear, for which a general solution has been
found recently during a study of semi-leptonic decays
\cite{Blokland:2004ye}.

The biggest  challenge is posed by the soft-soft diagrams such as
Fig.~\ref{fig:factors}(c).  Here  the imaginary part
of a genuine four-loop diagram is needed,
albeit now containing only a single
mass scale.  A related task was solved in a series of papers on the
muon and semi-leptonic $b\to u$ decays
\cite{vanRitbergen:1998yd,vanRitbergen:1999fi,vanRitbergen:1999gs}.
Those processes involve two non-interacting particles (neutrinos in
the case of the muon decay and leptons in the case of QCD
corrections to the $b\to u$ decay).  Our present problem is more
difficult because all particles can interact. For example,
all fermion lines in Fig.~\ref{fig:factors}(c) interact with
gluons.

For the purpose of this calculation  the algorithm proposed
in \cite{Laporta:2001dd} is  adopted.  Integration by parts
\cite{Tkachov:1981wb} generates identities through which
all needed integrals can be expresssed in terms of a few so-called master
integrals.  Some of them are the same as in Ref.
\cite{vanRitbergen:1999fi} and some new ones had to be determined,
as will be described elsewhere.  The large systems of linear
equations needed in this reduction procedure are solved using
symbolic manipulation software based on the BEAR package
\cite{bear}. Parts of the calculations were performed using the
package MINCER \cite{mincer} and programs for solving three-loop
diagrams developed in \cite{Blokland:2004nd}, using the computer
algebra program FORM \cite{Vermaseren:2000nd}.

As a result,
the two-gluon corrections
resulting from interactions between the quark lines are obtained in a fully
analytical form.
The result  can be presented as an enhancement of the hadronic decay width,
\ba
{\Gamma(b\to c\bar u d')\over 3\Gamma(b\to ce\bar \nu)}
=1+{\alpha_s\over \pi}
+\left({\alpha_s\over \pi}\right)^2
\cdot
4\left( L^2 + {15\over 8}L + \delta_1  + \delta_2\right) + \order{\alpha_s^3}.
\label{param}
\ea
Here $L \equiv \ln{M_W\over m_b}$ and $\delta_{1}$ describes the two-loop
corrections arising from interactions between the quarks $\bar u$ and $d'$.
For the purpose of this analysis, the value computed for the hadronic
decays of the lepton $\tau$ \cite{PDBook2004} is adopted (for a recent review of higher order
calculations for $\tau$ decays see \cite{Baikov:2005sw}),
\ba
\delta_1 \simeq 1.3.
\label{delta1}
\ea
The last correction, $\delta_2$, is the main result of the present research. It
arises from exchanges of two gluons between the $bc$ and $ud'$ lines and reads
\ba
\delta_2 =  - {9259\over 5832 }+ {17\over 18}\pi^2\ln 2
 - {5785\over 11664}\pi^2 + {13\over 1440}\pi^4 + {503\over 648}\zeta(3)
 \simeq 1.8.
\ea
The two non-logarithmic corrections, $\delta_{1,2}$, increase the non-leptonic
decay rate by $1.3\alpha_s^2$, or about $5$ to $8$ percent, if  $\alpha_s$ is varied between $0.2$ and $0.25$.
They increase the denominator of Eq.~(\ref{def}) and therefore lower the predicted semi-leptonic
branching ratio, by  0.35 to 0.6 percentage points.  Thus the theoretical prediction for $B_{\rm SL}$ goes down
from the previous ``reference point" of 11.5 percent in Eq.~(\ref{theory}) to about $11 $ percent,
certainly within one standard deviation from the  experimental determination, Eq.~(\ref{exp}).

In this study  all masses of the quarks in the final
state have been neglected.  This is likely a very good approximation for the decay $b\to
c\bar u d'$. For example, it is known how the coefficient of the
logarithm $L$ in Eq.~(\ref{param}) depends on $m_c$: it changes from
the massless limit value of $15/8=1.875$ to $1.79$ for the actual
charm mass \cite{Bagan:1994zd}.  It is less clear what the impact of
a non-zero $m_c$ is in the decay  $b\to c\bar c s'$.  In this
channel there are two massive charm quarks and they are moving slowly
in much of the available phase space.  This may greatly enhance
Coulomb-like strong interactions in the final state.  In the case of
a single gluon exchange between $\bar c$ and $s'$, the effect of
$m_c$ was found to increase the correction by more than a factor of
four \cite{Voloshin:1994sn}. Thus, it would be very valuable to
determine the impact of the charm mass on our correction $\delta_2$.
Such a study can be carried out using the technique developed in
this paper, but is technically even more challenging because of the
need to compute four-loop diagrams like Fig.~\ref{fig:factors}(c)
with higher powers of propagators.

The emerging agreement between the theoretical value and the
measurement of the semi-leptonic branching ratio confirms the
Standard Model description of the heavy quark dynamics.  Neither
large non-perturbative effects that could potentially arise from QCD
\cite{Bigi:1993fm} nor exotic contributions to the electroweak decay
mechanism are needed to bring theory and experiment into agreement.
Further improvements of perturbative calculations as well as
experimental studies of $B_{\rm SL}$ are very warranted. One may
hope to bring the comparison of theory and experiment to about $1 $ percent
level, and thus restrict or uncover ``new physics" contributions to
the heavy quark decays.

\emph{Acknowledgments:} We are grateful to M.~B.~Voloshin for
inspiring this project and many helpful discussions, and to D.~W.~Hertzog for
suggesting improvements of the manuscript.  Our
calculations were performed with the facilities of the Centre for
Symbolic Computation at the University of Alberta.


\end{document}